# Long-term stable poly(ionic liquid)/MWCNTs inks enable enhanced surface modification for electrooxidative detection and quantification of dsDNA

Larisa V. Sigolaeva[1,2]*, Tatiana V. Bulko[2], Maxim S. Kozin[2], Weiyi Zhang[3], Moritz Köhler[4], Iuliia Romanenko[4], Jiayin Yuan[3], Felix H. Schacher[4,5,6], Dmitry V. Pergushov[1], Victoria V. Shumyantseva[1,2]

[1]*Department of Chemistry, M.V. Lomonosov Moscow State University, 119991 Moscow, Russia*
[2]*V.N. Orekhovich Institute of Biomedical Chemistry, 119121 Moscow, Russia*
[3]*Department of Materials and Environmental Chemistry, Stockholm University, 10691 Stockholm, Sweden*
[4]*Institute of Organic Chemistry and Macromolecular Chemistry (IOMC), Friedrich-Schiller-University Jena, D-07743 Jena, Germany*
[5]*Jena Center for Soft Matter (JCSM), Friedrich-Schiller-University Jena, D-07743 Jena, Germany*
[6]*Center for Energy and Environmental Chemistry (CEEC), Friedrich-Schiller-University Jena, D-07743 Jena, Germany*

*corresponding author:
Larisa V. Sigolaeva, lsigolaeva@belozersky.msu.ru
Department of Chemistry, M.V. Lomonosov Moscow State University, Leninskie Gory 1/3, Moscow 119991, Russia, Tel.: +7 495 939 40 42

**Abstract**

This work demonstrates the use of imidazolium-based poly(ionic liquid)s (PILs) as efficient dispersants of multi-walled carbon nanotubes (MWCNTs). With these polymeric dispersants, highly stable fine dispersions of MWCNTs (inks) can be easily prepared in aqueous media and applied for rather simple but efficient surface modification of screen-printed electrodes (SPEs). Such a modification of SPEs remarkably increases the electroactive surface area and accelerates the electron transfer rate due to synergistic combination of specific features of MWCNTs such as strong adsorptive property and high specific surface with the advantages of PILs like ion conductivity and dispersability. We further show that the PIL/MWCNT-modified SPEs can be beneficially utilized for direct electrochemical analysis of double stranded DNA (dsDNA). Specifically, it is exemplified by the direct electrooxidation of guanine and adenine bases in salmon testes dsDNA chosen as a model system. The linear ranges for the determination of dsDNA correspond to 5–500 µg/mL for the oxidative peak of guanine and 0.5–50 µg/mL for the oxidative peak of adenine. This makes direct electrochemical dsDNA detection with the use of the easy-preparable PIL/MWCNT-modified SPEs strongly competing to currently applied spectral and fluorescent techniques. Furthermore, we show that the developed constructs are capable of sensing a single point mutation in the 12-bases single-stranded DNA fragments. Such detection is of high clinical significance in choosing an adequate anticancer treatment, where the electrochemical identification of the point mutation could offer time and cost benefits.

**Introduction**

Nucleotides, oligonucleotides, DNA, and RNA are known as markers of many pathological states. The presence of mutations in circulating tumor DNA (ctDNA) as well as in RNA or microRNA (miRNA) and their amount can serve as diagnostic biomarkers for various oncologic diseases, predictive markers for analyzing the response to medical treatment, and/or as markers of disease progression [1]. Thereby, alteration of DNA length (DNA fragmentation) is one of the accepted markers of programmed cell death (apoptosis) [2-4]. In that regard, the analysis of modified heterocyclic bases (e.g. methylation profiles) is prominent for epigenetic studies [5] as well as for the detection of point mutations [6]. Actively developing gene editing technologies are promising for applied medicine from the viewpoint of genome correction [7]. In the context of the important functional role of nucleic acids as well as both their natural and modified components, different analytical techniques, e.g., high-performance liquid chromatography, capillary electrophoresis, laser-induced fluorescence detection, chemoluminescence, gas chromatography, or mass-spectrometry are actively developing nowadays [8, 9].

The analysis using polymerase chain reaction (PCR) is the most highly-demanded method to reveal nucleotide sequences in DNA, while amplification allows for the detection of traces [10]. At the same time, PCR-based diagnostics are time-consuming and laborious, also requiring expensive reagents and equipment. New methods for analytical quantification of DNA and/or miRNA in biological samples are highly required, as concentrations of the nucleic acids can be used as a marker of pathological states. However, these concentrations are often either not known or only the extrapolated values coming from the PCR assay can be obtained. Absorbance and fluorescence (based on the Quant-iT$^{TM}$ reagent) techniques allow for calculating the total content of nucleotide material in the sample without selective identification of individual heterocyclic bases [11, 12]. Yet still it is so far impossible to analyze mononucleotide substitutions (single nucleotide polymorphism) by these approaches [13].

Electroanalysis of heterocyclic bases, nucleotides, DNA and RNA (direct electrochemistry of DNA bases) is based on the appearance of individual oxidative peaks of purines: guanine (G) and adenine (A) (typically at < 1.0 V potentials) and pyrimidines: cytosine (C) and thymine (T) (at higher positive potentials), following a pH-dependent mechanism of electrooxidation [14]. For heterocyclic bases taken at equal concentrations, however, the currents corresponding to electrooxidation of pyrimidines (T and C) are lower compared to the ones for electrooxidation of purines (A and G) [8]. Moreover, the oxidation of pyrimidines T and C occurs at high positive potentials (typically higher than +1.2 V) [14]. Both these points make electrochemical detection of pyrimidines with disposable screen-printed electrodes (SPEs) rather challenging as SPEs have a limited potential window applied for measurements. At the same time, such electrodes are being widely used for experiments with biological samples (plasma, blood serum, whole blood, urine, saliva, tissues, cells, etc.) [15]. Consequently, the electrochemical detection of oxidative changes occurring in DNA centers around mainly on the appearance of purine base oxidation peaks. Quantitative detection of purines A and G also allows for evaluating the number of pyrimidines as A = T and G = C, according to the Chargaff's rule.

There are some additional unbiased difficulties in quantitative electrochemical DNA assay. It worth noting that an electrochemical signal coming from dsDNA was lower than that of ssDNA of the same size because of less number of A- and G- residues accessible for the electrooxidation in rigid dsDNA than flexible ssDNA. From this viewpoint, the electrochemical assay of ssDNA is expected to be more sensitive. Nevertheless, direct electrochemical detection of native dsDNA could be more informative from biomedical point of view as it allows avoiding additional sample treatment (thermal denaturation for dsDNA, dsDNA ↔ ssDNA transition [16] or, if necessary, a complete hydrolysis of dsDNA to heterocyclic nucleobases under harsh conditions [8]) that can lead to distortion of results and incorrect DNA analysis [12, 17]. Hence, the premodification of electrodes has primary importance for dsDNA detection and quantification.

To enhance electroanalysis sensitivity, chemical modification of working electrodes with nanostructured carbon materials such as carbon nanotubes (CNTs) and, in particular, multi-walled CNTs (MWCNTs), graphene, ordered mesoporous carbon, or carbon nanofibers, is widely used [17-19]. Besides, commercial availability MWCNTs imparts the modified electrodes a number of beneficial properties, e.g., increased surface area, improved conductivity, and an expanded potential window.

Drop-casting seems to be the most simple and easiest way of electrode surface modification by CNTs. For uniform coverage of a working electrode, a highly homogeneous dispersion of MWCNTs with good conductivity and without (or at least minimal) structural damage is required. This is, however, rather challenging because of poor dispersability of such carbon nanomaterials in most typical solvents. Highly destructive MWCNT functionalization via harsh oxidation in the concentrated mixture of nitric and sulfuric acids is nowadays replaced by more safe and soft ultrasound treatment either in water or in organic solvents (dimethylformamide, acetone, isopropanol, ethanol, toluene, N-methyl-2-pyrrolidone, cyclodextrines). Quite often, surfactants and polymers (sodium dodecylsulfate, Nafion, chitosan) are also used to improve the dispersability of MWCNTs and to enhance colloidal stability of their dispersions [20].

Ionic liquids (ILs), also well-known as green electrolytes and solvents, possess a number of valuable properties such as negligible vapor pressure, high ionic conductivity, physicochemical and thermal stability, ion-exchange properties, and catalytic activity. The application of ILs is widely explored in electrochemistry due to their broad electrochemical window and high ionic conductivity as well as the ability to couple with carbon materials to produce IL-carbon nanomaterial hybrids [21, 22]. Electroanalysis of DNA and/or heterocyclic bases has also been reported for electrodes prepared from mixtures of carbon nanomaterial (graphite powder, graphene, or CNTs) with ILs [23-27].

Imidazolium-based poly(ionic liquid)s (PILs) represent an even more innovative approach to obtain carbon nanomaterial composites [28-30] due to π–cation interactions between the imidazolium ring and the surface of the carbon nanomaterials [31]. Apart from dispersing, the PIL/CNT nanocomposites can be prepared by an alternative synthetic procedure consisting of initially dispersing CNTs in an IL, followed by bulk polymerization of the latter [32]. A similar strategy was used for the synthesis of PIL/CNT hydrogels and organogels [33]. Polymerization of imidazolium-based ILs on MWCNT surfaces also produces a highly homogeneous suspension of CNTs with good conductivity that is beneficial for the fabrication of electrochemical sensors [34].

A number of biosensors were reported, that used PILs as a polymeric matrix for glucose oxidase [35, 36] or hemoglobin [37] immobilization or to produce PIL-based nanocomposite electrodes with PIL-wrapped CNTs, Pt nanoparticles, or graphene [38-40]. Another PIL-based support for ultrafine Pt nanoparticles was described for non-enzymatic hydrogen peroxide detection [41]. However, the use of PILs itself or dispersions of PIL-stabilized carbon nanomaterials for direct electrochemical analysis of dsDNA has not been reported yet.

We highlight herein the preparation of long-term stable colloidal dispersions of MWCNTs by imidazolium-based cationic PILs: poly(1-ethyl-3-vinylimidazolium bromide) and poly(1-butyl-3-vinylimidazolium bromide). Hence, the poly(ionic liquid)s are a key component of the nanocomposite coatings, with the polymer acting as a polymeric binder of the carbon nanomaterial. The following easy modification of SPEs with PIL/MWCNT nanocomposite inks results in rather easy preparable and universal sensor surface which advances we demonstrate for direct electrochemical analysis of dsDNA.

**Materials and Methods**

**Materials**

Poly(1-ethyl-3-vinylimidazolium bromide) (PIL-Et) and poly(1-butyl-3-vinylimidazolium bromide) (PIL-But) were synthesized similar to the method described in [42]. Briefly, using IL-Et as an example: 10.38 g of IL-Et (50.8 mmol), 30 mg of AIBN (0.183 mmol), 70 mL of ethanol were loaded into a 250 mL reactor. The mixture was deoxygenated 3 times by a freeze-pump-thaw procedure. The reactor was then refilled with nitrogen and placed in an oil bath at 70 °C for 24 hours. The mixture was then exhaustively dialyzed against water for 3 days and freeze-dried from water. $^1$H-NMR spectra of synthesized PIL-Et and PIL-But are shown in Supporting Information (Figures S1, S2). $MWCNT_1$ with an outer diameter of 10–15 nm, an inner diameter of 2–6 nm, and a length of 0.1–10 μm and $MWCNT_2$ with a mean diameter of 9.5 nm, a length of 1 μm were obtained from Sigma-Aldrich. $K_3[Fe(CN)_6]$ and $K_4[Fe(CN)_6]$ were obtained from Reakhim (Russia). dsDNA of salmon sperm (testes) with water content of ≤10% and guanosine triphosphate (GTP) were purchased from Sigma-Aldrich. Oligonucleotides, 5'-AAACCCGCCCGG-3 and 5'- AAACCCGACCGG-3, were synthesized by Eurogene Co. (Russia). Phosphate buffer solutions with NaCl (100 mM potassium phosphate with 50 mM NaCl) with specified pHs were prepared by mixing stock solutions of 100 mM $KH_2PO_4$/50 mM NaCl and 100 mM $K_2HPO_4$/50 mM NaCl to the desired pH-value. All other chemicals were of analytical grade and were used without further purification. All aqueous solutions were prepared using Milli-Q water (18.2 MΩ cm) purified with a Milli-Q water purification system by Millipore.

**Methods**

**Dispersing of Carbon Nanomaterials.** An aqueous solution of each of the PILs (PIL-Et or PIL-But) was prepared in a concentration of 3 g/L. Then, a portion of $MWCNT_1$ or $MWCNT_2$ was added to the aqueous solution of the corresponding PIL in such a way that each 1 mL of the PIL solution contains 1 mg of the carbon nanomaterial. Such prepared mixtures were treated in Ultrasound Desintegrator SONOPULS HD 4100, 100W at 20% power in pulse regime.

**Preparation of Electrochemical Sensors.** Three-pronged screen-printed electrodes (SPEs) purchased from Color Electronics (Russia, http://www.colorel.ru) were used for the electrode preparation. They consist of a round graphite working electrode (2 mm in diameter) surrounded by a graphite ringed auxiliary counter-electrode and an Ag/AgCl reference electrode. For the preparation of modified electrodes, 2 μL of the corresponding MWCNT dispersion in a PIL was dropped onto the working electrode and incubated for 15 min at 37 °C until complete drying. For further incorporation of the analyte (GTP or dsDNA), 2 μL of the solution with a specified concentration in 100 mM potassium phosphate with 50 mM NaCl (pH 7.4) were dropped onto a 2 mm area of the freshly prepared modified electrode and incubated for 15 min at 37 °C until complete drying. Such prepared modified SPEs were stored refrigerated at +4 °C until measurement on the same day.

**Scanning Electron Microscopy (SEM).** SEM was performed with a Sigma VP Field Emission Scanning Electron Microscope (Carl-Zeiss AG, Germany) operating at 5 to 10 kV using an InLens detector. Previously, the samples were coated with gold (5 nm) using a SCD005 sputtering device BAL-TEC (Balzers, Liechtenstein). For the cross-section view, two small opposite cuts were done in the edges of SPE close to working electrode area with a help of scissors. Then, SPEs were frozen in liquid nitrogen, broken and the working electrode region was peeled off from polymeric support with the help of a pincer. The peeled off pieces were subjects for cross-view analysis. The SEM micrographs in Supporting Information were obtained using a Hitachi S-5500 Scanning Electron Microscope operating at 25 kV.

**Transmission Electron Microscopy (TEM).** For TEM measurements, copper grids were rendered hydrophilic by Ar plasma cleaning for 2 min (Diener Electronics). 10 µL of the respective sample solution were applied onto the grid and excess sample was blotted with a filter paper. TEM images were acquired with a 200 kV FEI Tecnai G2 20 equipped with a 4k × 4k Eagle HS CCD and a 1k × 1k Olympus MegaView camera for overview images.

**Electrochemical Measurements.** Cyclic voltammetry (CV) and differential pulse voltammetry (DPV) measurements were performed using an Autolab PGSTAT12 potentiostat/galvanostat (Metrohm Autolab, the Netherlands) equipped with the GPES software (version 4.9.7). All electrochemical experiments were carried out at room temperature in 100 mM potassium phosphate with 50 mM NaCl of pH 7.4. CV experiments were carried out in a 1 mL electrochemical cell by potential sweeping from an initial potential of -300 mV to an end-point potential of +800 mV at different scan rates in a range of 10-100 mV/s. DPV experiments were carried out in a 60 µL drop applied onto the SPE to cover all the three electrodes. The following experimental DPV parameters were used: potential range of 0.2–1.2 V, pulse amplitude of 0.025 V, potential step of 0.005 V, pulse duration of 50 ms, modulation amplitude of 0.05 V. All potentials were referred to the Ag/AgCl reference electrode. For the DPV experiments with accumulation, the accumulation potential of 0.4 V and the accumulation time of 5 min were applied prior the measurements.

**Results and discussion**

**1. Preparation and Characterization of PIL/MWCNT Dispersions**

From a macroscopic point of view, MWCNTs represent powders that are hardly wettable by water and most organic solvents. This makes them difficult to be handled for the modification of electrode surfaces in general and dispersing of MWCNTs in a suitable solvent (the most preferable is water) is a challenging task. To overcome this problem, we employed PIL-Et and PIL-But (Figure 1) as stabilizers.

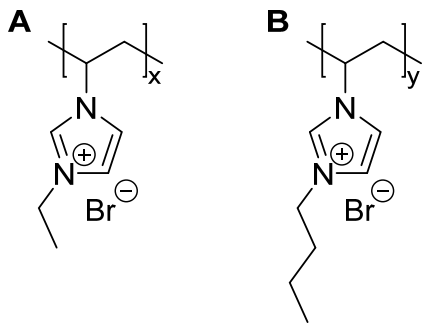

**Figure 1.** Chemical structures of (A) poly(1-ethyl-3-vinylimidazolium bromide) (PIL-Et) and (B) poly(1-butyl-3-vinylimidazolium bromide) (PIL-But).

In spite of being permanent and strong polyelectrolytes, PILs are known to be soluble in many types of organic solvents with different polarity [28]. In general, PILs can adopt physical properties across a large spectrum. Hence, we reasonably suppose that their unique properties make them rather good dispersants for carbon nanomaterials such as MWCNTs. Moreover, we expect that the macromolecular structure of PILs can provide integrity of a MWCNT layer upon the electrode modification, with PIL acting as a binder.

Two different types of commercial MWCNTs were tested, i.e., $MWCNT_1$ with an outer diameter of 10–15 nm, an inner diameter of 2–6 nm, and a length of 0.1–10 μm and $MWCNT_2$ with a mean outer diameter of 9.5 nm and a length of 1 μm. It is remarkable that the simple mixing of pristine MWCNTs with aqueous solutions of PIL-Et or PIL-But does not result in the formation of dispersions (Figure 2A). A non-covalent functionalization of the MWCNTs with PILs was carried out by ultrasonication of the respective mixtures. After ultrasonication, rather uniform black suspensions can be observed (Figure 2B, samples 1, 2, 5, 6). Once being ultrasonicated, the prepared dispersions of MWCNTs in aqueous solutions of PIL-Et or PIL-But do not precipitate and stay colloidally-stable at room temperature. At least by now, no precipitation has been detected over more than 3 years (See Supporting Information, Figure S3). Contrary to this, similar ultrasonication of of the same MWCNTs (also at 1 g/L) in chloroform results in poor dispersions containing rather large pieces/flakes of non-dispersed material clearly visible by eye (Figures 2B, 2C, samples 4, 8). Any attempts to ultrasonicate MWCNTs in water failed, leading to very large aggregates (Figure 2B, 2C, samples 3, 7). Repeated or prolonged ultrasonication does not improve the situation.

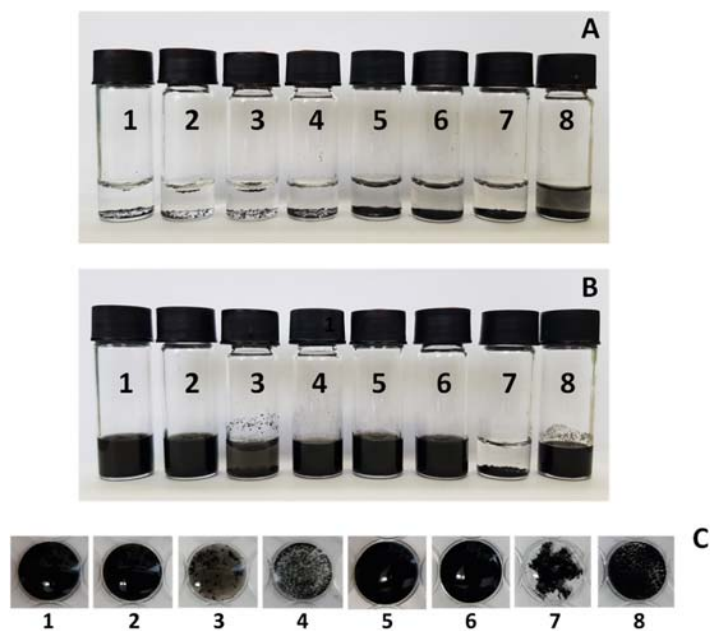

**Figure 2.** Photographs of mixtures of 1 mg of MWCNTs$_{1,2}$ with 3 g/L aqueous solutions of PIL-Et or PIL-But: initial mixtures before any treatment (A), mixtures a day after ultrasonication treatment (B), the appearance of dispersions in a thin layer (C). (1) PIL-Et/MWCNT$_1$; (2) PIL-But/MWCNT$_1$; (3) H$_2$O/MWCNT$_1$; (4) chloroform/MWCNT$_1$; (5) PIL-Et/MWCNT$_2$; (6) PIL-But/MWCNT$_2$; (7) H$_2$O/MWCNT$_2$; (1) chloroform/MWCNT$_2$.

The quality of the prepared PIL/MWCNT dispersions was further evaluated by TEM (Figure 3). From TEM images, one can see a dominant presence of individual MWCNTs, although one cannot solidly confirm the presence of a wrapping layer at the MWCNT surface, while others report the presence of a very thin layer of PIL wrapped SWCNTs, with a thickness the PIL layer of several nanometers [43]. The agglomerates (albeit non-precipitating) are also observable for dispersions of MWCNT$_1$ (the sample with the increased MWCNT length) in aqueous solutions of PILs. The dispersions of shorter MWCNT$_2$ look more homogeneous. It seems to be that the length of MWCNT is an important factor determining their dispersability.

To get an impression how the prepared PIL/MWCNT nanocomposites are distributed across the electrode surface, we drop-casted 2 μL drops of each dispersion onto an active electrode area of SPEs. After drying, the surface morphology of the modified SPEs was examined by SEM both from the top as well as using cross-sectional analysis. The obtained SEM images demonstrate a complete coverage of the electrode surface by a condensed layer of MWCNTs as can be seen from exemplified images shown in Figure 4 for MWCNTs$_{1,2}$ dispersed in aqueous solutions of PIL-Et or PIL-But. The top view of the surface of SPEs modified by PIL-Et/MWCNT$_1$ or PIL-But/MWCNT$_1$ dispersions shows rather rough coating. This is obviously due to the presence of residual bundles of MWCNTs. More uniform coatings were achieved, as expected, in the case of SPE modified by PIL-Et/MWCNT$_2$ or PIL-But/MWCNT$_2$ dispersions. These observations are in line with TEM imaging (Figure 3). The thicknesses of the modifying MWCNT layers were obtained from the corresponding cross-section views (Figure 4, Insets) and were found to be in a range of 2.0–4.0 μm for PIL-Et/MWCNT$_1$, 1.4–1.7 μm for PIL-But/MWCNT$_1$, 0.6–0.8 μm for PIL-Et/MWCNT$_2$, and 1.5–2.0 μm for PIL-But/MWCNT$_2$. It is worth noting that similar surface modifications of SPEs with chloroform dispersions of MWCNTs$_{1,2}$ led to less efficient and less homogeneous (discontinuous) coatings. In both cases, the areas were covered by large aggregates of MWCNTs and the areas of non-covered bare SPE surface can be clearly observed in the corresponding SEM images (Supporting Information, Figure S4) with equal probability. Thus, we can conclude that the uniform coating of SPEs after modification using PIL/MWCNT dispersions suggests mutual compatibility between the PILs and the MWCNTs. This is expected to provide advantages for electrochemical measurements.

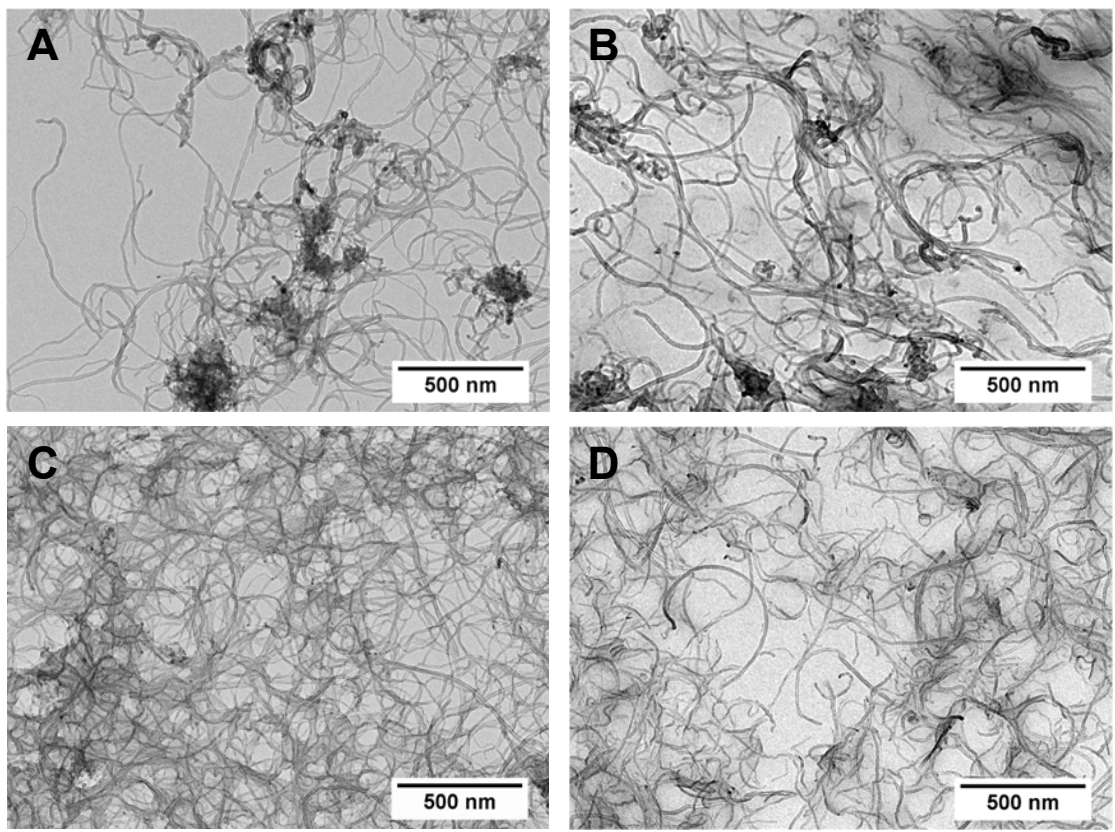

**Figure 3.** TEM micrographs of the PIL-Et/MWCNT$_1$ (A); PIL-But/MWCNT$_1$ (B); PIL-Et/MWCNT$_2$ (C); and PIL-But/MWCNT$_2$ (D). Dispersions were prepared by ultrasonication of 1 g/L of MWCNTs in aqueous solutions of PILs with the concentration of 3 g/L.

Further, uniform dispersions can be prepared by ultrasonication of MWCNTs in aqueous solutions of PILs, where individual MWCNTs seem to be wrapped by macromolecules of PIL-Et or PIL-But. We assume that strong positive charges and high charge density of PIL chains (Figure 1) considerably (or even decisively) contribute to the stabilization of MWCNTs in aqueous media. It is worth noting that a polymer layer wrapping MWCNTs seems to be very thin as it is not visible by TEM/SEM techniques.

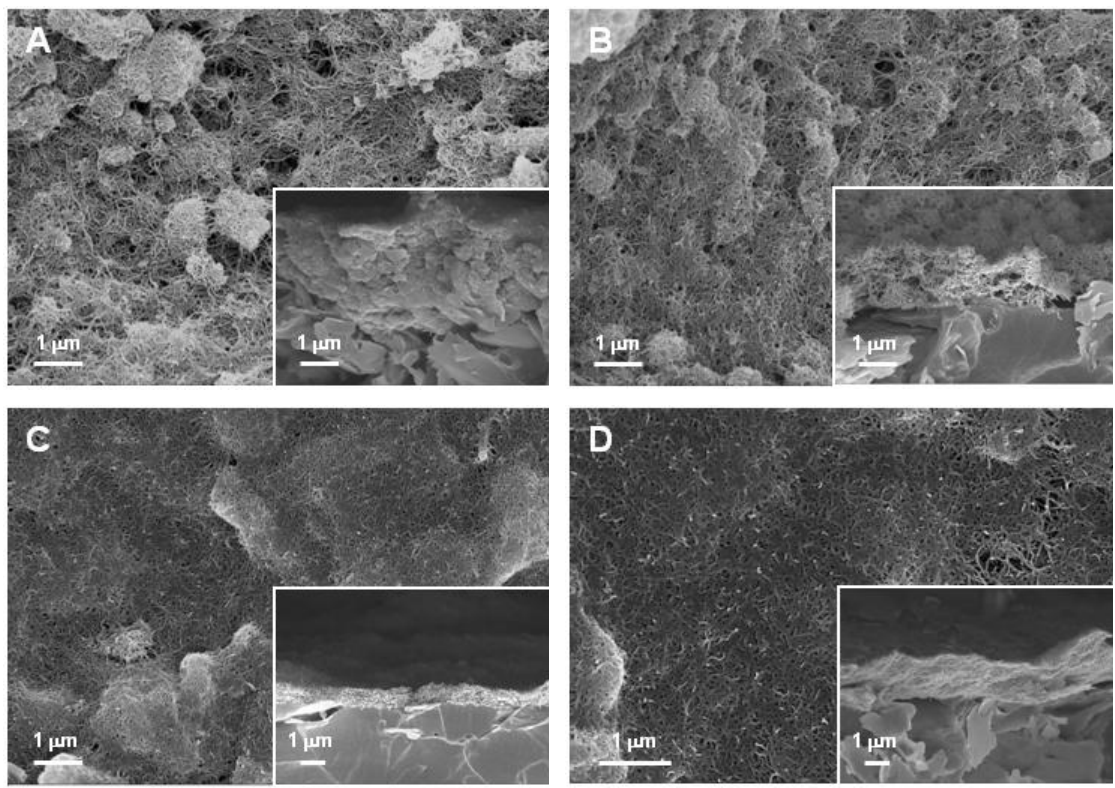

**Figure 4.** SEM micrographs of the PIL-Et/MWCNT$_1$ (A), PIL-But/MWCNT$_1$ (B), PIL-Et/MWCNT$_2$ (C), and PIL-But/MWCNT$_2$ (D) dispersions drop-casted onto the surface of SPEs. Insets show corresponding cross-section views.

## 2. Electrochemical Characterization of the SPEs Modified by PIL/MWCNT Dispersions

The voltammetric behavior of the modified SPEs was studied by CV of typically used electroactive species of $K_3[Fe(CN)_6]/K_4[Fe(CH)_6]$ as redox probe at different scan rates (Figure 5, more CV data can be found in Supporting Information). The values of $E_{red}$, $E_{ox}$, $\Delta E$, $E_{1/2}$, $I_{red}$, $I_{ox}$ were determined at a scan rate of 50 mV/s and were compared for all modified SPEs. The results of electrochemical characterization are summarized in Table 1. The data given in Figure 5 and Table 1 shows that the naked SPE and the SPE modified by PILs both show poor redox peaks with the peak separation of >230 mV and low current responses (Table 1, Supporting Information, Figures S5-S10). Alternatively, a pair of well-defined reversible redox peaks with a narrow peak-to-peak separation of <122 mV was observed in case of the SPEs modified with the PIL/MWCNT dispersions. This implies that more-reversible redox performance of $[Fe(CN)_6]^{3-/4-}$ occurs when MWCNTs are used. Furthermore, a considerable increase in the current response is found as well, thereby demonstrating enhanced electron transfer properties, an increasing mass transfer, and more reversibility for the SPEs modified by a nanocomposite based on the PIL/MWCNT dispersions.

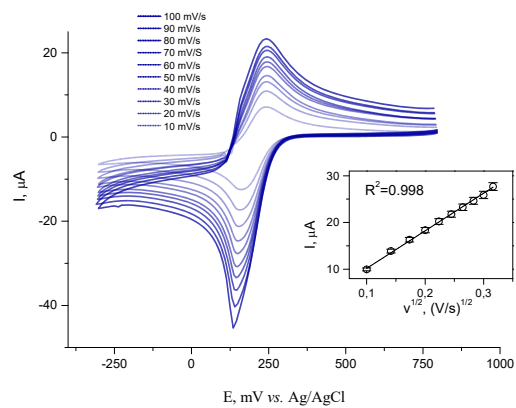

**Figure 5.** Typical CVs for the SPE/(PIL-Et/MWCNT$_1$). The measurements were carried out in 5 mM of K$_3$[Fe(CN)$_6$] at ambient temperature in potential range from -300 mV to +800 mV (*vs* Ag/AgCl) at scan rates in a range of 10–100 mV/s. Inset: The dependence of peak current I$_p$ on the square root of scan rate in the range of 10–100 mV/s.

As seen from the inset of Figure 5, the anodic/cathodic peak currents exhibit a linear relationship with the square root of the scan rate in the range of 10–100 mV/s, providing evidence for a chemically reversible redox process. The electroactive surface area, A (cm$^2$), of the electrode can be defined using the Randles–Sevcik equation:

$$Ip = 2.69 * 10^5 * A * \sqrt{D} * \sqrt{v} * C * n^{3/2} \qquad (1),$$

where $I_p$ is the current maximum in amps, $A$ is the electroactive surface area in cm$^2$, $D$ is the diffusion coefficient in cm$^2$/s, $v$ is the scan rate in V/s, , $C$ is the concentration in mol/cm$^3$, and $n$ is the number of electrons transferred in the redox event.

By substituting the known values of $D = 7.6 \times 10^{-6}$ cm$^2$/s [44], $n = 1$, and $C = 0.005$ M for the used K$_3$[Fe(CN)$_6$]/K$_4$[Fe(CH)$_6$] redox probe into the equation (1), the surface areas of each electrode modification can easily be extracted from the slope with the dependence of $I_p$ vs $v^{1/2}$. The calculated values of $A$ are also summarized in Table 1. The results convincingly demonstrate that the modification of the SPEs by the PIL/MWCNT dispersions results in a considerable increase in electroactive surface area. Interestingly, the dispersions prepared with the shorter MWCNT$_2$ appear to be even more effective modifying agents for the SPEs than samples treated with comparable dispersions of MWCNT$_1$. These data are in line with TEM and SEM experiments. The reported results suggest a beneficial application of the PIL/MWCNT dispersions for the modification of SPEs that seems to considerably facilitate the electron transfer between redox probe and electroactive electrode surface.

**Table 1.** Comparison of the electrochemical characteristics of the modified SPEs[*].

| SPE modification | $E_{red}$, mV | $E_{ox}$, mV | $\Delta E$, mV | $E_{1/2}$, mV | $I_{red}$, µA | $I_{ox}$, µA | Electroactive surface area, A, cm$^2$ | Electroactive surface area relative to control (naked SPE) |
|---|---|---|---|---|---|---|---|---|
| Naked SPE | -87 ± 8 | 421 ± 41 | 499 ± 52 | 167 ± 18 | 6.2 ± 0.5 | 5.0 ± 0.5 | 0.0024 (7.7%[**]) | 1.0 |
| SPE/PIL-Et | 70 ± 7 | 299 ± 32 | 229 ± 32 | 185 ± 19 | 8.9 ± 0.8 | 6.0 ± 0.6 | 0.0027 (8.7%) | 1.1 |
| SPE/PIL-But | 7 ± 8 | 470 ± 50 | 463 ± 51 | 238 ± 25 | 28.0 ± 5.0 | 24.0 ± 2.4 | 0.0052 (16.6%) | 2.2 |
| SPE/(PIL-Et/MWCNT$_1$) | 151 ± 14 | 243 ± 25 | 92 ± 9 | 197 ± 21 | 24.4 ± 5.0 | 20.2 ± 2.0 | 0.0077 (24.4%) | 3.2 |
| SPE/(PIL-But/MWCNT$_1$) | 123 ± 12 | 241 ± 24 | 118 ± 11 | 182 ± 18 | 26.7 ± 5.0 | 19.8 ± 2.0 | 0.0091 (28.9%) | 3.8 |
| SPE/(PIL-Et/MWCNT$_2$) | 146 ± 15 | 234 ± 22 | 88 ± 9 | 190 ± 19 | 34.1 ± 6.0 | 24.4 ± 2.1 | 0.0117 (37.4%) | 4.9 |
| SPE/(PIL-But/MWCNT$_2$) | 112 ± 10 | 234 ± 24 | 122 ± 13 | 173 ± 18 | 34.5 ± 6.0 | 24.0 ± 2.2 | 0.0120 (38.4%) | 5.0 |

[*] Values of $E_{red}$, $E_{ox}$, $I_{red}$, $I_{ox}$ are given for a potential rate of 50 mV/s. All potentials are given *vs.* Ag/AgCl reference electrode.
[**] Calculated with respect to geometric electrode area of 0.0314 cm$^2$.

## 3. Characterization and Quantification of Electrochemical dsDNA Assay

The sample of dsDNA of salmon tests was used as a model source of dsDNA and was directly deposited onto surfaces of the above-mentioned SPEs pre-modified by the PIL/MWCNT dispersions. Experimentally, 2 µL drops of a dsDNA solution were deposited onto the active area of the electrode, followed by 15 min incubation at +37 °C for dsDNA adsorption and then accompanied by drying of the drop. Further, DPV was employed to follow the direct electrooxidation of dsDNA and its subsequent quantification. One typical example of voltammograms obtained by DPV for the SPE/(PIL-Et/MWCNT$_1$) is given in Figure 6. In the absence of dsDNA, no obvious peak was seen during the potential scan. After the addition of dsDNA, two concentration-dependent irreversible oxidation peaks appear at about +0.6 V and +0.9 V. This corresponds to the known pH-dependent mechanism of electrochemical oxidation of G- and A-residues, which are present in dsDNA [8]. The rather large potential gap of 0.3 V between these peaks points out that the electrochemical responses of G- and A-residues do not interfere. The known oxidation mechanism of DNA involves the irreversible oxidation of such residues according to the scheme shown in Figure 6 [14].

The DPV responses for 3 mg/mL dsDNA were determined and compared for all kinds of the modified SPEs (Table 2). As one can see, a naked SPE or the SPEs modified by PILs only show one rather weak peak at about +950 mV corresponding to the oxidation of A-residues, with the current responses in the range of several nA. Alternatively, a huge increase in the DPV response was found for all the SPE modifications with the PIL/MWCNT dispersions, resulting in two well-defined oxidative peaks of both G-and A-residues. The comparison of the data presented in Table 2 allows one to conclude that the highest oxidative peaks were found for the SPE/(PIL-Et/MWCNT$_2$) and the SPE/(PIL-But/MWCNT$_2$) modifications. Among them, the SPE/(PIL-But/MWCNT$_2$) modification appears to be the most preferable as it shows more reproducible response with variation coefficients in the range of 17–25%. Taking this fact into account, we used the SPE/(PIL-But/MWCNT$_2$) modification for all further experiments.

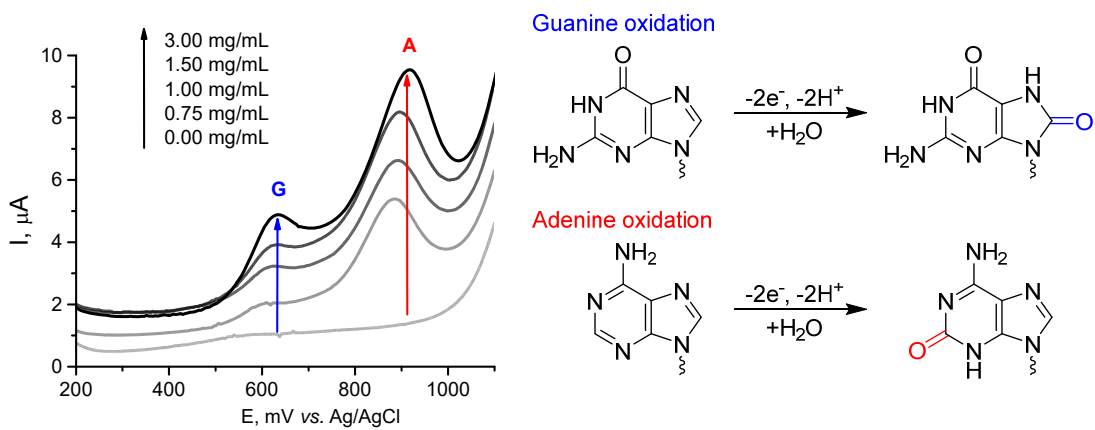

**Figure 6.** The typical voltammograms obtained by DPVs for the dsDNA oxidation at the SPE/(PIL-Et/MWCNT$_2$) together with corresponding mechanisms of electrochemical oxidation of guanine (G) and adenine (A) residues.

**Table 2.** Comparative electrochemical characteristics of dsDNA oxidation on the modified SPEs.

| Electrode modification | Guanine (G) residue | | Adenine (A) residue | |
|---|---|---|---|---|
| | V, mV | I, nA (CV, %) | V, mV | I, nA (CV, %) |
| Naked SPE | - | - | 935 ± 17 | 4.9 ± 3.1 |
| SPE/PIL-Et | - | - | 957 ± 5 | 3.1 ± 1.3 |
| SPE/PIL-But | - | - | 952 ± 22 | 1.0 ± 0.8 |
| SPE/(PIL-Et/MWCNT$_1$) | 636 ± 30 | 521 ± 214 (41) | 898 ± 17 | 1371 ± 729 (53) |
| SPE/(PIL-But/MWCNT$_1$) | 634 ± 13 | 250 ± 42 (17) | 918 ± 14 | 1270 ± 467 (37) |
| SPE/(PIL-Et/MWCNT$_2$) | 599 ± 10 | 922 ± 276 (30) | 881 ± 14 | 4260 ± 1430 (34) |
| SPE/(PIL-But/MWCNT$_2$) | 610 ± 11 | 632 ± 108 (17) | 890 ± 13 | 4200 ± 1065 (25) |

The efficiency of the dsDNA adsorption onto the electrode surface directly influences the overall performance of the electrochemical detection of dsDNA. A pre-concentration of dsDNA at the electrode surface *via* electrostimulation of its adsorption could result in an improved sensitivity of the DNA assay. The factors that have a direct influence on the dsDNA adsorption onto the SPEs are the accumulation potential and the time of accumulation. Hence, both these factors were studied and optimized using DPV, following the oxidative peak of GTP (Supporting Information, Figures S11, S12). Therefore, the accumulation potential of +400 mV *vs.* Ag/AgCl and the accumulation time of 5 min were selected as the optimum accumulation parameters, thus resulting in a further 2.5–3 fold increase in the dsDNA responses.

Under optimized conditions, we examined oxidative peak currents $I_G$ and $I_A$ with regard to the concentration of a dsDNA solution deposited onto the SPE/(PIL-But/MWCNT$_2$/) constructs. The changes of $I_G$ and $I_A$ against the dsDNA concentration are shown in Figure 7, with the initial linear parts of the calibration curves being given in the corresponding insets. The analytical parameters that were extracted from the corresponding calibration curves, like linearity, sensitivity, detection limit, are summarized in Table 3. The relative standard deviation (SD) was assessed for at least three measurements (n ≥ 3) at each concentration level and was found to be within 5%. This demonstrates that the quantification of dsDNA with the SPE/(PIL-But/MWCNT$_2$) constructs is reproducible. Thus, a rather wide concentration range of 0.5–500 μg/mL is accessible for the quantitative dsDNA detection using our herein described SPE/(PIL-But/MWCNT$_2$) constructs. This makes the dsDNA DPV detection using such constructs a strong alternative to currently applied spectral and fluorescent techniques.

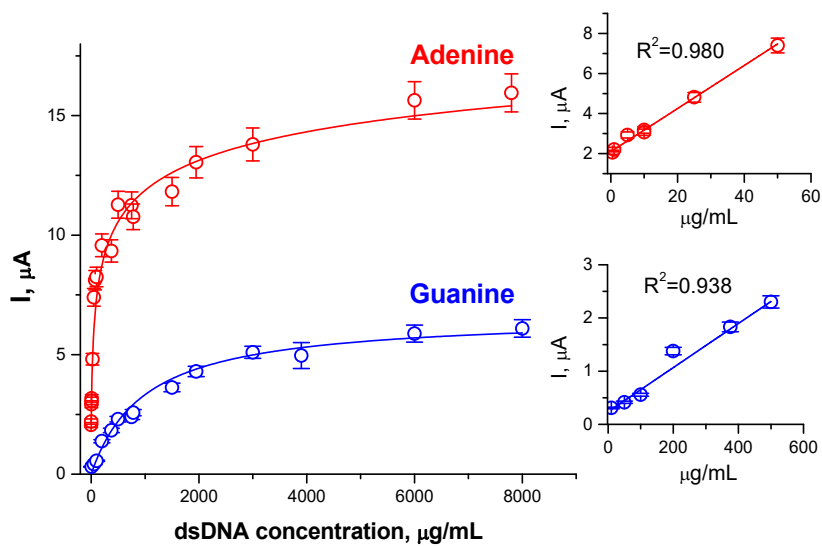

**Figure 7.** Oxidative peak current $I_G$ of guanine (G) residues (blue) and $I_A$ of adenine (A) residues (red) versus the concentration of dsDNA for the SPE/(PIL-But/MWCNT$_2$). Technique: DPV coupled with sample accumulation by stripping voltammetry, accumulation potential of +400 mV, accumulation time of 5 min, potential range of 200–1200 mV, potential step of 5 mV, modulation amplitude of 50 mV.

**Table 3.** The electroanalytical parameters of quantitative dsDNA assay with PIL/MWCNT-modified SPEs.

|  | Guanine (G) residue | Adenine (A) residue |
|---|---|---|
| $E_{ox}$, mV | 600 ± 5 | 850 ± 5 |
| Sensitivity, $\mu A/(\mu g \times mL^{-1} \times cm^2)$ | 0.14 | 3.36 |
| Liearity range, µg/mL | 5–500 | 0.5–50 |
| Detection limit, µg/mL | 5 | 0.5 |
| Equation for linear regression[*)] | $I_G = 0.0042[dsDNA] + 0.2673$ | $I_A = 0.1057[dsDNA] + 2.1324$ |

*) $I_G$, $I_A$ correspond to the oxidative currents (peak heights) for G- and A-residues, respectively;
[dsDNA] is a concentration of dsDNA solution in mg/mL that was deposited onto the electrode active area.

As the oxidative peak heights are proportional to the content of the corresponding A- or G-residues in the nucleotide chain, the direct electrooxidation can be used for the detection of point mutations or so-called single nucleotide polymorphism in short DNA fragments (oligonucleotides). To demonstrate this, a pair of oligonucleotides was used, which differs only in one nucleotide residue:

Oligonucleotide 1: 5'- AAACCCG**A**CCGG -3',

Oligonucleotide 2: 5'- AAACCCG**C**CCGG–3'.

The oligonucleotide sequence and the position of the mutation were specifically synthesized as full analogues of the growth factor receptor (EGFR) 21 fragment. Non-small cell lung carcinoma is associated with the epidermal growth factor receptor (EGFR) exon 21 L858R and, as known, the disease has a rather poor prognosis [6]. The exon 21 L858R detection has high clinical significance in choosing adequate treatment, where the electrochemical identification of such a point mutation could be rather useful.

Figure 8 clearly shows a difference in $I_A$-values found for these oligonucleotides: the oxidative current for A-residues was found to be about 5 µA and 3 µA for oligonucleotide 1 and 2, respectively, due to the different content of A-residues, while the oxidative current for G-residues was similar for both oligonucleotides indicating no difference in the G-residues content. With this experiment, thus, we demonstrate a potential ability of the developed SPE/(PIL-But/MWCNT$_2$) construct for the future detection of point mutations in short DNA fragments.

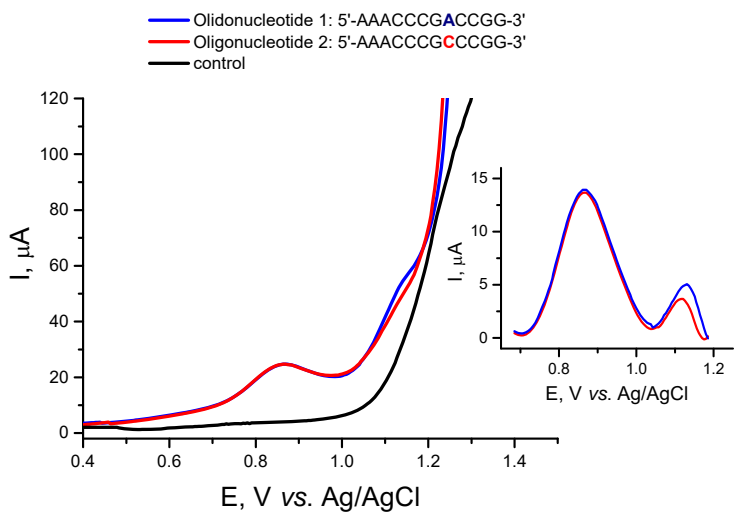

**Figure 8.** The DPV plot for oligonucleotide 1: 5'-AAACCCG**A**CCGG-3' (blue) or oligonucleotide 2: 5'-AAACCCG**C**CCGG-3' (red) oxidation at the SPE/(PIL-But/MWCNT$_2$) after deposition of 2 µL of the 50 µM solution of the corresponding oligonucleotide in 100 mM potassium phosphate with 50 mM NaCl of pH 7.4. Measurement conditions: DPV coupled with sample accumulation by stripping voltammetry, accumulation potential of +0.4 V, accumulation time of 5 min, potential range of 0.2–1.2 V, potential step of 0.005 V, modulation amplitude of 0.05 V.

**Conclusion**

In this work, highly stable fine dispersions of MWCNTs in aqueous solutions of imidazolium-based PILs, poly(1-ethyl-3-vinylimidazolium bromide) and poly(1-butyl-3-vinylimidazolium bromide), were prepared and characterized. The modification of SPEs with well-defined PIL-based aqueous dispersions of the MWCNTs was found to remarkably accelerate the electron transfer rate, to increase the electroactive surface area, and to impart high conductivity to the naked (bare) SPE. The PIL/MWCNT-modified SPEs were for the first time utilized for examining the direct electrooxidation of G- and A-bases in salmon testes dsDNA used as a model system. Two well-defined irreversible oxidation peaks corresponding to the oxidation of G- and A-residues were taken for the quantitative detection of dsDNA after its direct adsorption on the surface of the modified SPEs.

We assume that the specific characteristics of MWCNTs, such as strong adsorptive property and high specific surface area, combined with the advantages of PILs, such as high dispersing power, contribute to the efficient adsorption of dsDNA onto the SPE and therefore to sensitivity of the dsDNA assay. Such a synergistic effect of MWCNTs combined with an ionic amphiphilic (micelle-forming) diblock copolymer was already reported by us previously with respect to the quantitative electrochemical detection of myoglobin [45]. The herein described easy-preparable PIL/MWCNT-modified SPEs compared to other electrochemical setups intended for the dsDNA quantification exhibit a similar or better sensitivity for the electrochemical oxidation of dsDNA [26, 46]. The linear ranges for the determination of dsDNA correspond to 5–500 μg/mL (for G) and 0.5–50 μg/mL (for A). In addition, we demonstrate that the developed PIL/MWCNT-modified SPEs are able to sense a point mutation in the 12-bases single-stranded DNA fragments. Such detection is of high clinical significance in choosing an adequate anticancer treatment, where the electrochemical identification of the point mutation could offer time and cost benefits. We are sure that our strategy can be readily applied for the development of convenient and robust biosensors for numerous practical applications.

**Acknowledgments:** This research was supported by the Russian Science Foundation (RSF, project no. 18-44-04011) and the Deutsche Forschungsgemeinschaft (DFG, SCHA1640/18-1) within a joint RSF-DFG grant. The German side also gratefully acknowledges support from the SFB 1278 (PolyTarget, Project C03) and SFB-TRR 234 (CataLight, Project B05). The SEM/TEM facilities of the Jena Center for Soft Matter (JCSM) were established with a grant from the DFG and the European Regional Development Fund (ERDF). The authors are grateful to Dr. Sergey L. Kanashenko and Dr. Yurii D. Ivanov for SEM imaging of SPEs modified by MWCNT dispersions in chloroform. J. Yuan thanks the ERC Starting Grant 639720 NAPOLI and the Wallenberg Academy Fellow program (KAW 2017.0166) from the Knut and Alice Wallenbergs Foundation in Sweden.

**Conflicts of Interest:** The authors declare no conflict of interest.

**Supplementary Material:** Figure S1: $^1$H-NMR spectrum of PIL-Et and the assignment of all chemical shifts to the chemical structure. Figure S2: $^1$H-NMR spectrum of PIL-But and the assignment of all chemical shifts to the chemical structure. Figure S3: Appearance of the dispersions of MWCNTs upon a time. Figure S4: SEM images of $MWCNT_1$ (A, B) and $MWCNT_2$ (C, D) dispersions in chloroform drop-casted onto the surface of SPEs. Figures S5-S10: The CVs for different SPE modifications obtained at different scan rates. Figure S11: The effect of the accumulation potential on the DPV responses of GTP. Figure S12: The effect of the accumulation time on the DPV responses of GTP.

**References**

[1] S. Nikolaev, L. Lemmens, T. Koessler, J.L. Blouin, T. Nouspikel, Circulating tumoral DNA: Preanalytical validation and quality control in a diagnostic laboratory, Analytical Biochemistry 542 (2018) 34-39.

[2] M. Hasanzadeh, N. Shadjou, M. de la Guardia, Early stage diagnosis of programmed cell death (apoptosis) using electroanalysis: Nanomaterial and methods overview, Trac-Trends in Analytical Chemistry 93 (2017) 199-211.

[3] J. Yin, P. Miao, Apoptosis Evaluation by Electrochemical Techniques, Chemistry-an Asian Journal 11(5) (2016) 632-641.

[4] I.M. Huffnagle, A. Joyner, B. Rumble, S. Hysa, D. Rudel, E.G. Hvastkovs, Dual Electrochemical and Physiological Apoptosis Assay Detection of in Vivo Generated Nickel Chloride Induced DNA Damage in Caenorhabditis elegans, Analytical Chemistry 86(16) (2014) 8418-8424.

[5] I. Sanjuan, A.N. Martin-Gomez, J. Graham, N. Hernandez-Ibanez, C. Banks, T. Thiemann, J. Iniesta, The electrochemistry of 5-halocytosines at carbon based electrodes towards epigenetic sensing, Electrochimica Acta 282 (2018) 459-468.

[6] Y. Shoja, A. Kermanpur, F. Karimzadeh, Diagnosis of EGFR exon21 L858R point mutation as lung cancer biomarker by electrochemical DNA biosensor based on reduced graphene oxide/functionalized ordered mesoporous carbon/Ni-oxytetracycline metallopolymer nanoparticles modified pencil graphite electrode, Biosensors & Bioelectronics 113 (2018) 108-115.

[7] R.O. Bak, N. Gomez-Ospina, M.H. Porteus, Gene Editing on Center Stage, Trends in Genetics 34(8) (2018) 600-611.

[8] S. Ren, H. Wang, H.Y. Zhang, L.Q. Yu, M.J. Li, M. Li, Direct electrocatalytic and simultaneous determination of. purine and pyrimidine DNA bases using novel mesoporous carbon fibers as electrocatalyst, Journal of Electroanalytical Chemistry 750 (2015) 65-73.

[9] L. Svorc, K. Kalcher, Modification-free electrochemical approach for sensitive monitoring of purine DNA bases: Simultaneous determination of guanine and adenine in biological samples using boron-doped diamond electrode, Sensors and Actuators B-Chemical 194 (2014) 332-342.

[10] S. Carinelli, M. Kuhnemund, M. Nilsson, M.I. Pividori, Yoctomole electrochemical genosensing of Ebola virus cDNA by rolling circle and circle to circle amplification, Biosensors & Bioelectronics 93 (2017) 65-71.

[11] O. Aybastier, S. Dawbaa, C. Demir, O. Akgun, E. Ulukaya, F. Ari, Quantification of DNA damage products by gas chromatography tandem mass spectrometry in lung cell lines and prevention effect of thyme antioxidants on oxidative induced DNA damage, Mutation Research-Fundamental and Molecular Mechanisms of Mutagenesis 808 (2018) 1-9.

[12] O. Tangvarasittichai, W. Jaiwang, S. Tangvarasittichai, The Plasma DNA Concentration as a Potential Breast Cancer Screening Marker, Indian Journal of Clinical Biochemistry 30(1) (2015) 55-58.

[13] K. Chang, S.L. Deng, M. Chen, Novel biosensing methodologies for improving the detection of single nucleotide polymorphism, Biosensors & Bioelectronics 66 (2015) 297-307.

[14] P.N. Bartlett, Bioelectrochemistry: Fundamentals, Experimental Techniques and Applications, Wiley ed., Wiley, England, 2008.

[15] F. Arduini, L. Micheli, D. Moscone, G. Palleschi, S. Piermarini, F. Ricci, G. Volpe, Electrochemical biosensors based on nanomodified screen-printed electrodes: Recent applications in clinical analysis, Trac-Trends in Analytical Chemistry 79 (2016) 114-126.

[16] Y. Fan, K.J. Huang, D.J. Niu, C.P. Yang, Q.S. Jing, $TiO_2$-graphene nanocomposite for electrochemical sensing of adenine and guanine, Electrochimica Acta 56(12) (2011) 4685-4690.

[17] C. Corro, T. Hejhal, C. Poyet, T. Sulser, T. Hermanns, T. Winder, G. Prager, P.J. Wild, I. Frew, H. Moch, M. Rechsteiner, Detecting circulating tumor DNA in renal cancer: An open challenge, Experimental and Molecular Pathology 102(2) (2017) 255-261.

[18] A. Abi, Z. Mohammadpour, X.L. Zuo, A. Safavi, Nucleic acid-based electrochemical nanobiosensors, Biosensors & Bioelectronics 102 (2018) 479-489.

[19] S. Gupta, C.N. Murthy, C.R. Prabha, Recent advances in carbon nanotube based electrochemical biosensors, International Journal of Biological Macromolecules 108 (2018) 687-703.

[20] C. Hu, S. Hu, Carbon nanotube-based electrochemical sensors: principles and applications in biomedical systems, Journal of Sensors 2009 (2009) 1-40.

[21] A. Abo-Hamad, M.A. AlSaadi, M. Hayyan, I. Juneidi, M.A. Hashim, Ionic Liquid-Carbon Nanomaterial Hybrids for Electrochemical Sensor Applications: a Review, Electrochimica Acta 193 (2016) 321-343.

[22] S.G. Zhang, Q.H. Zhang, Y. Zhang, Z.J. Chen, M. Watanabe, Y.Q. Deng, Beyond solvents and electrolytes: Ionic liquids-based advanced functional materials, Progress in Materials Science 77 (2016) 80-124.

[23] W. Sun, Y.Z. Li, Y.Y. Duan, K. Jiao, Direct electrocatalytic oxidation of adenine and guanine on carbon ionic liquid electrode and the simultaneous determination, Biosensors & Bioelectronics 24(4) (2008) 988-993.

[24] H.W. Gao, Y.Y. Duan, L. Xu, W. Sun, Electrochemical Oxidation of 2'-deoxyguanosine-5'-triphosphate on Ionic Liquid Modified Carbon Paste Microelectrode and its Sensitive Detection, Croatica Chemica Acta 84(1) (2011) 33-38.

[25] M. Arvand, A. Niazi, R.M. Mazhabi, P. Biparva, Direct electrochemistry of adenine on multiwalled carbon nanotube-ionic liquid composite film modified carbon paste electrode and its determination in DNA, Journal of Molecular Liquids 173 (2012) 1-7.

[26] J.F. Ping, S.P. Ru, X. Luo, K. Fan, J. Wu, Y.B. Ying, Direct electrochemistry of double strand DNA on ionic liquid modified screen-printed graphite electrode, Electrochimica Acta 56(11) (2011) 4154-4158.

[27] J. Yang, T. Yang, S.F. Hou, A sensor based on polyaniline nanofibers/ionic liquid-functionalized carbon nanotubes composite for electrocatalytic oxidation of guanine, Journal of the Iranian Chemical Society 13(9) (2016) 1611-1615.

[28] J. Yuan, D. Mecerreyes, M. Antonietti, Poly(ionic liquid)s: An update, Progress in Polymer Science 38(7) (2013) 1009-1036.

[29] Y.J. Men, X.H. Li, M. Antonietti, J. Yuan, Poly(tetrabutylphosphonium 4-styrenesulfonate): a poly(ionic liquid) stabilizer for graphene being multi-responsive, Polymer Chemistry 3(4) (2012) 871-873.

[30] S. Soll, M. Antonietti, J. Yuan, Double Stimuli-Responsive Copolymer Stabilizers for Multiwalled Carbon Nanotubes, ACS Macro Letters 1(1) (2012) 84-87.

[31] T. Fukushima, A. Kosaka, Y. Ishimura, T. Yamamoto, T. Takigawa, N. Ishii, T. Aida, Molecular ordering of organic molten salts triggered by single-walled carbon nanotubes, Science 300(5628) (2003) 2072-2074.

[32] T. Fukushima, A. Kosaka, Y. Yamamoto, T. Aimiya, S. Notazawa, T. Takigawa, T. Inabe, T. Aida, Dramatic effect of dispersed carbon nanotubes on the mechanical and electroconductive properties of polymers derived from ionic liquids, Small 2(4) (2006) 554-560.

[33] S.H. Hong, T.T. Tung, L.K.H. Trang, T.Y. Kim, K.S. Suh, Preparation of single-walled carbon nanotube (SWNT) gel composites using poly(ionic liquids), Colloid and Polymer Science 288(9) (2010) 1013-1018.

[34] X.L. Wang, K.Y. Zheng, X. Feng, C.H. Xu, W.B. Song, Polymeric ionic liquid functionalized MWCNTs as efficient electrochemical interface for biomolecules simultaneous determination, Sensors and Actuators B-Chemical 219 (2015) 361-369.

[35] S. Lee, B.S. Ringstrand, D.A. Stone, M.A. Firestone, Electrochemical Activity of Glucose Oxidase on a Poly(ionic liquid)-Au Nanoparticle Composite, Acs Applied Materials & Interfaces 4(5) (2012) 2311-2317.

[36] M.S.P. Lopez, D. Mecerreyes, E. Lopez-Cabarcos, B. Lopez-Ruiz, Amperometric glucose biosensor based on polymerized ionic liquid microparticles, Biosensors & Bioelectronics 21(12) (2006) 2320-2328.

[37] Q. Zhang, X. Lv, Y. Qiao, L. Zhang, D.L. Liu, W. Zhang, G.X. Han, X.M. Song, Direct Electrochemistry and Electrocatalysis of Hemoglobin Immobilized in a Polymeric Ionic Liquid Film, Electroanalysis 22(9) (2010) 1000-1004.

[38] X.C. Chu, B.H. Wu, C.H. Xiao, X.H. Zhang, J.H. Chen, A new amperometric glucose biosensor based on platinum nanoparticles/polymerized ionic liquid-carbon nanotubes nanocomposites, Electrochimica Acta 55(8) (2010) 2848-2852.

[39] C.H. Xiao, X.C. Chu, B.H. Wu, H.L. Pang, X.H. Zhang, J.H. Chen, Polymerized ionic liquid-wrapped carbon nanotubes: The promising composites for direct electrochemistry and biosensing of redox protein, Talanta 80(5) (2010) 1719-1724.

[40] Q. Zhang, S.Y. Wu, L. Zhang, J. Lu, F. Verproot, Y. Liu, Z.Q. Xing, J.H. Li, X.M. Song, Fabrication of polymeric ionic liquid/graphene nanocomposite for glucose oxidase immobilization and direct electrochemistry, Biosensors & Bioelectronics 26(5) (2011) 2632-2637.

[41] X.J. Bo, J. Bai, B. Qi, L.P. Guo, Ultra-fine Pt nanoparticles supported on ionic liquid polymer-functionalized ordered mesoporous carbons for nonenzymatic hydrogen peroxide detection, Biosensors & Bioelectronics 28(1) (2011) 77-83.

[42] D. Kuzmicz, P. Coupillaud, Y. Men, J. Vignolle, G. Vendraminetto, M. Ambrogi, D. Taton, J. Yuan, Functional mesoporous poly(ionic liquid)-based copolymer monoliths: From synthesis to catalysis and microporous carbon production, Polymer 55(16) (2014) 3423-3430.

[43] T. Kim, T.T. Tung, T. Lee, J. Kim, K.S. Suh, Poly(ionic liquid)-Mediated Hybridization of Single-Walled Carbon Nanotubes and Conducting Polymers, Chemistry-an Asian Journal 5(2) (2010) 256-260.

[44] H.C. Chen, C.C. Chang, K.H. Yang, F.D. Mai, C.L. Tseng, L.Y. Chen, B.J. Hwang, Y.C. Liu, Polypyrrole electrode with a greater electroactive surface electrochemically polymerized in plasmon-activated water, Journal of the Taiwan Institute of Chemical Engineers 82 (2018) 252-260.

[45] V.V. Shumyantseva, L.V. Sigolaeva, L.E. Agafonova, T.V. Bulko, D.V. Pergushov, F.H. Schacher, A.I. Archakov, Facilitated biosensing via direct electron transfer of myoglobin integrated into diblock copolymer/multi-walled carbon nanotube nanocomposites, Journal of Materials Chemistry B 3(27) (2015) 5467-5477.

[46] P. Canete-Rosales, A. Alvarez-Lueje, S. Bollo, Ethylendiamine-functionalized multi-walled carbon nanotubes prevent cationic dispersant use in the electrochemical detection of dsDNA, Sensors and Actuators B-Chemical 191 (2014) 688-694.

Supporting Information

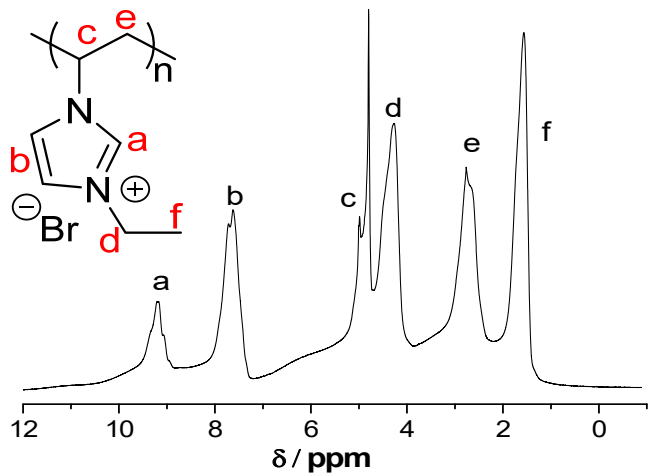

**Figure S1.** $^1$H-NMR spectrum of PIL-Et and the assignment of all chemical shifts to its chemical structure.

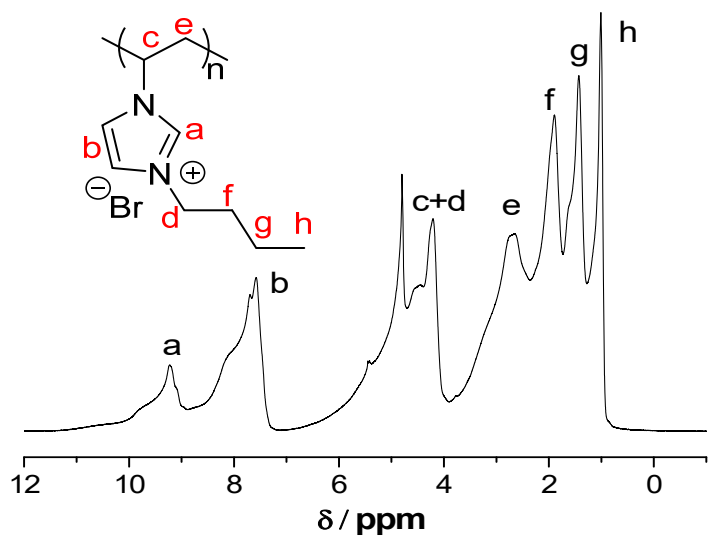

**Figure S2.** $^1$H-NMR spectrum of PIL-But and the assignment of all chemical shifts to its chemical structure.

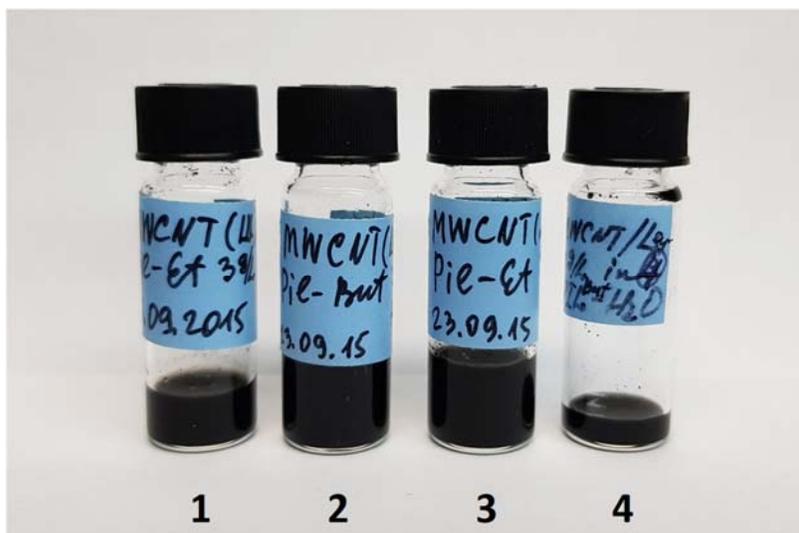

**Figure S3.** The appearance of the PIL/MWCNT dispersions upon a time (1) PIL-Et/MWCNT$_1$ (2) PIL-But/MWCNT$_1$; (3) PIL-Et/MWCNT$_2$; (4) PIL-But/MWCNT$_2$.

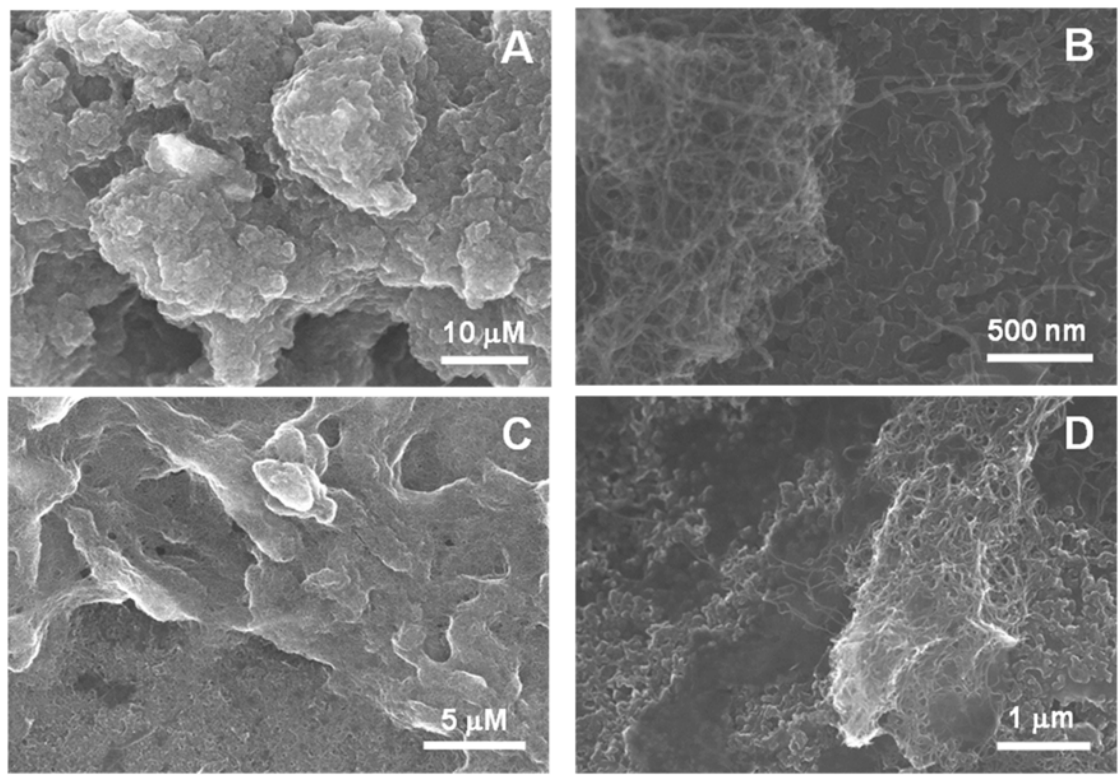

**Figure S4.** SEM micrographs of chloroform dispersions of MWCNT$_1$ (A, B) and MWCNT$_2$ (C, D) drop-casted onto the surface of the SPEs.

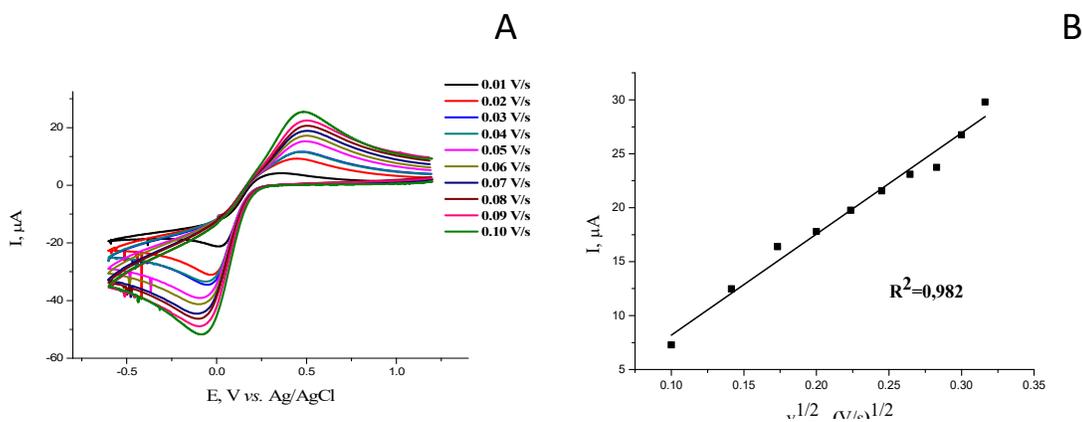

**Figure S5.** (A): The typical CVs for SPE/(PIL-But/MWCNT$_1$). The measurements were carried out in 5 mM of K$_3$Fe(CN)$_6$ at ambient temperature in potential range from -600 mV to +1200 mV (*vs.* Ag/AgCl), at scan rates in a range of 10–100 мВ/с. (B): The dependence of peak current I$_p$ on the square root of scan rate in the range of 10–100 mV/s.

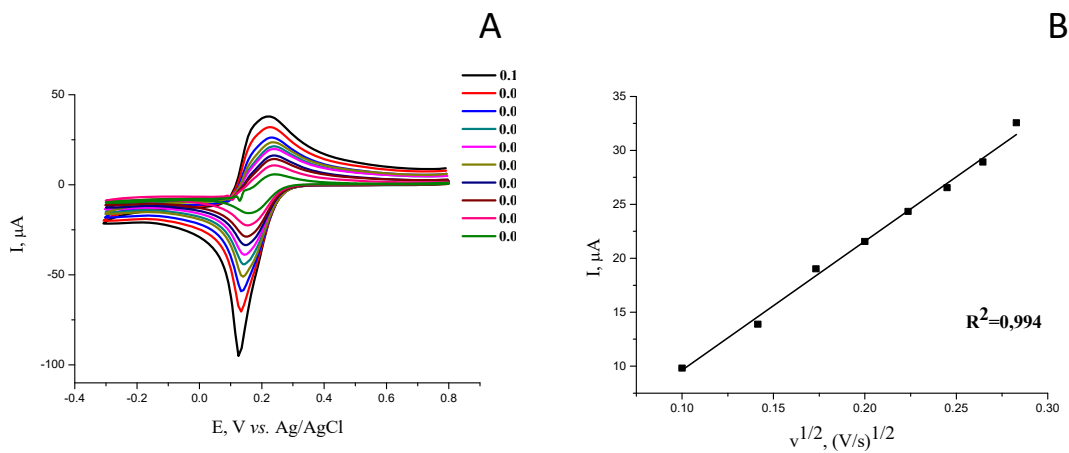

**Figure S6.** (A): The typical CVs for SPE/(PIL-Et/MWCNT$_2$). The measurements were carried out in 5 mM of K$_3$Fe(CN)$_6$ at ambient temperature in potential range from -300 mV to +800 mV (*vs.* Ag/AgCl), at scan rates in a range of 10–100 мВ/с. (B): The dependence of peak current I$_p$ on the square root of scan rate in the range of 10–100 mV/s.

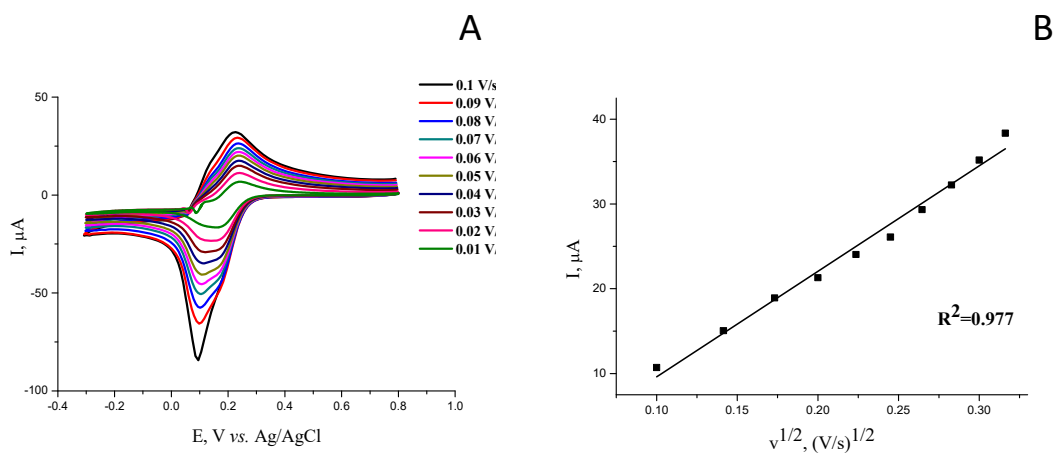

**Figure S7.** (A): The typical CVs for SPE/(PIL-But/MWCNT$_2$). The measurements were carried out in 5 mM of K$_3$Fe(CN)$_6$ at ambient temperature in potential range from -300 mV to +800 mV (*vs* Ag/AgCl), at scan rates in a range of 10–100 мВ/с. (B): The dependence of peak current I$_p$ on the square root of scan rate in the range of 10–100 mV/s.

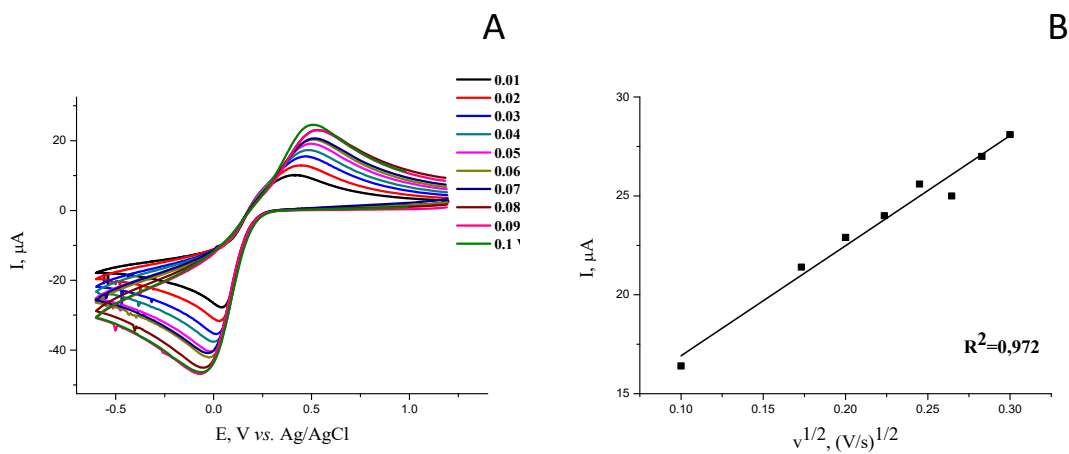

**Figure S8.** (A): The typical CVs for SPE/PIL-But. The measurements were carried out in 5 mM of $K_3Fe(CN)_6$ at ambient temperature in potential range from -600 mV to +1200 mV (*vs.* Ag/AgCl), at scan rates in a range of 10–100 мВ/с. (B): The dependence of peak current $I_p$ on the square root of scan rate in the range of 10–100 mV/s.

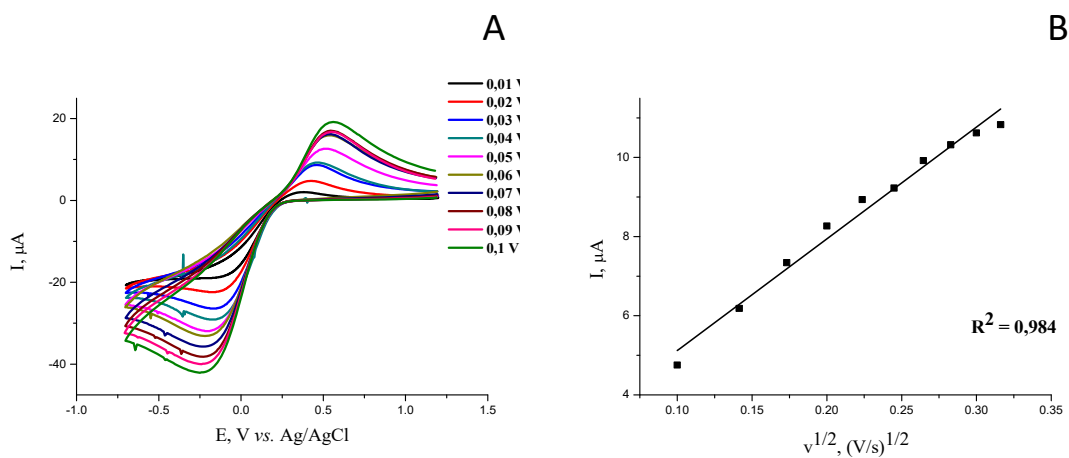

**Figure S9.** (A): The typical CVs for SPE/PIL-Et. The measurements were carried out in 5 mM of $K_3Fe(CN)_6$ at ambient temperature in potential range from -600 mV to +1200 mV (*vs.* Ag/AgCl), at scan rates in a range of 10–100 мВ/с. (B): The dependence of peak current $I_p$ on the square root of scan rate in the range of 10–100 mV/s.

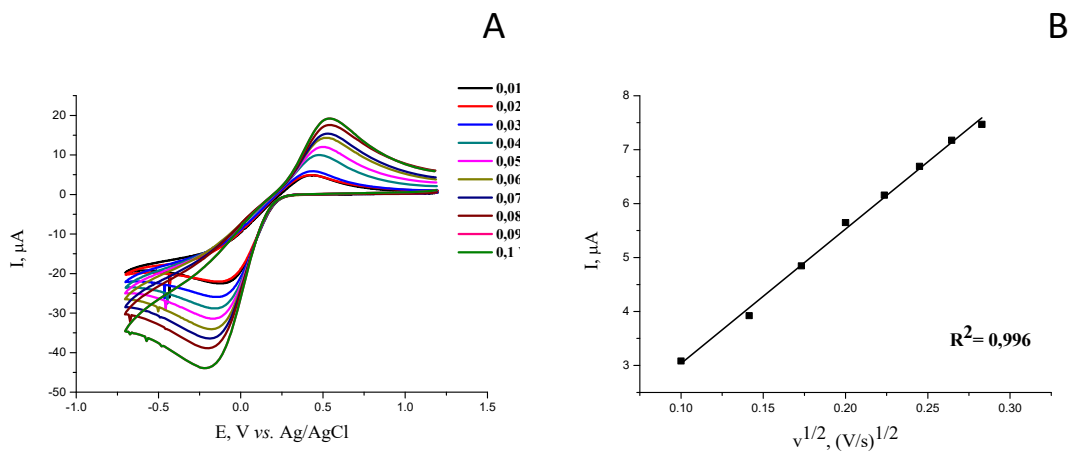

**Figure S10.** (A): The typical CVs for the naked SPE. The measurements were carried out in 5 mM of $K_3Fe(CN)_6$ at ambient temperature in potential range from -600 mV to +1200 mV (*vs.* Ag/AgCl), at scan rates in a range of 10–100 мВ/с. (B): The dependence of peak current $I_p$ on the square root of scan rate in the range of 10–100 mV/s.

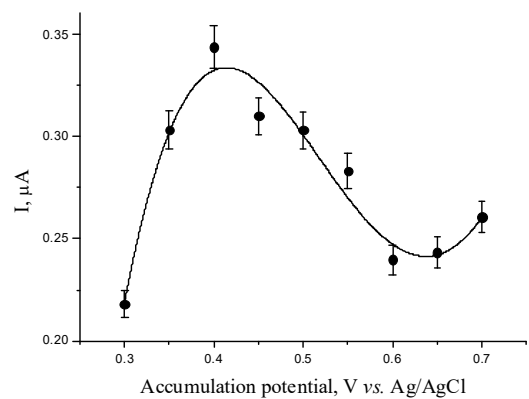

**Figure S11.** The effect of accumulation potential on the oxidative peak current after deposition of 2 μL of the 50 mM solution of GTP on a surface of SPE/(PIL-But/MWCNT$_2$) electrode. Conditions: linear voltammetry at scan rate 100 mV/s.

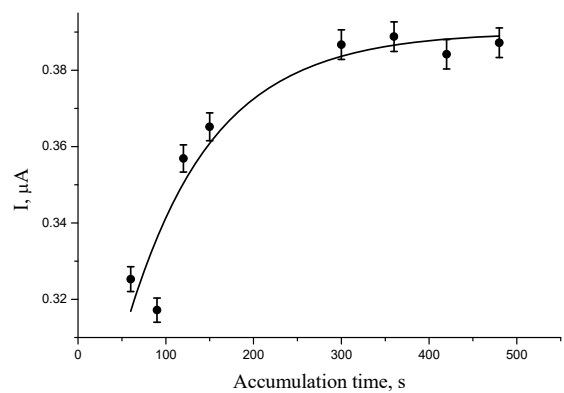

**Figure S12.** The effect of accumulation time on the oxidative peak current after deposition of 2 μL of the 50 mM solution of GTP on a surface of SPE/(PIL-But/MWCNT$_2$) electrode. Conditions: linear voltammetry at scan rate 100 mV/s.